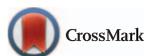

PAPER

# Optical nuclear electric resonance in LiNa: selective addressing of nuclear spins through pulsed lasers

Johannes K Krondorfer , Matthias Diez and Andreas W Hauser
Institute of Experimental Physics, Graz University of Technology, Petersgasse 16, A-8010 Graz, Austria

E-mail: andreas.w.hauser@gmail.com

## Abstract
Optical nuclear electric resonance (ONER), a recently proposed protocol for nuclear spin manipulation in atomic systems via short laser pulses with MHz repetition rate, exploits the coupling between the nuclear quadrupole moment of a suitable atom and the periodic modulations of the electric field gradient generated by an optically stimulated electronic excitation. In this theory paper, we extend the scope of ONER from atomic to molecular systems and show that molecular vibrations do not interfere with our protocol. Exploring the diatomic molecule LiNa as a first benchmark system, our investigation showcases the robustness with respect to molecular vibration, and the ability to address and manipulate each of the two nuclear spins independently, simply by adjusting the repetition rate of a pulsed laser. Our findings suggest that it might be possible to shift complicated spin manipulation tasks required for quantum computing into the time domain by pulse-duration encoded laser signals.

## 1. Introduction

In recent years, quantum technologies have undergone a remarkable transformation, opening new avenues for solving complex problems that were previously considered insurmountable for classical computing. Among the various approaches to quantum computing, the manipulation of nuclear spins for quantum information processing has emerged as an area of immense promise [1–6], offering the advantage of comparably large coherence times. In larger ensembles, manipulation and readout are well-established [7–9], but a spatially selective addressing of single spins via magnetic fields still remains a limiting factor. The control and confinement of electric fields, on the other hand, is an industrial standard. A suitable handle to access spin states via electric signals is provided by the nuclear quadrupole moment. Coherent quadrupole coupling, a possibility suggested many decades ago [10], has just recently been confirmed experimentally 'by accident' for a high-spin nucleus interacting with a faulty connection wire acting as a radio frequency antenna [3]. Besides nuclear electric resonance (NER), also nuclear acoustic resonance (NAR), a coherent coupling of acoustic phonons and nuclear spin states mediated by nuclear quadrupole interaction, has been observed experimentally [3, 11, 12].

For the sake of an improved spatial resolution, and taking current advancements in nano-optics and laser technology into consideration, we have recently suggested the application of pulsed lasers in the uv/visible regime as a potential tool for nuclear spin manipulation [13]. As a third paradigm aside NER and NAR, our protocol, optical nuclear electric resonance (ONER), is accessing nuclear spin states via a pulsed modulation of the electric field gradient (EFG) at the position of the nucleus of interest. This is achieved by a periodic excitation of a suitable atomic or molecular system into an electronically excited state with a pronounced change in the EFG. In this case, a carefully chosen repetition rate of the optical pulses allows for a coherent control of nuclear spin states. In our introductive paper, we have chosen the $^1P^o \leftarrow {}^1S$ transition in atomic $^9$Be for a first theoretical study. In the current paper, we extend our formalism towards molecules, choosing LiNa as a benchmark system. This heteronuclear alkali dimer has been studied in fields such as ultracold reactivity and collision dynamics, photoassociative spectroscopy of cold atoms, and quantum computing. Within the context of ONER, our primary objectives are twofold. First, we show that the protocol is not hindered by molecular vibration. Second, we demonstrate that it allows for individual addressing of nuclear spins, either at the Li or the Na nucleus, through a pulsed laser of same wavelength but different repetition rate.

We discuss the computational prerequisites for the derivation of all model parameters from electronic structure theory for the benchmark molecular system LiNa. This is followed by an evaluation of the potential energy surfaces (PES) of the $^1X\Sigma^+$ ground state as well as the $^1A\Sigma^+$ electronically excited state via electronic structure theory. Sigma states have been picked for the sake of zero angular momentum in the electronic wavefunction, eliminating any potential complications through additional coupling mechanisms. With relevant molecular parameters, PES scans, and EFG data in place, we discuss the ONER protocol for the LiNa molecule and provide spectroscopic predictions for a future validation by experiment.

## 2. Theoretical background

Atomic nuclei consist of protons and neutrons and therefore exhibit a charge distribution. The latter has a vanishing dipole moment with respect to the center of charge of the nucleus, but might possess a non-vanishing quadrupole moment, which is related to the nuclear spin $I$ and interacts with the electric field gradient (EFG), the second derivative of the electric potential, at the position of the nucleus. When placed into an external magnetic field $B_0$, an additional Zeeman term appears, and the total Hamiltonian reads

$$H = H_B + H_Q = -\gamma_n B_0 I_z + I_\mu Q_{\mu\nu} I_\nu, \tag{1}$$

where we have introduced the nuclear quadrupole interaction (NQI) tensor $Q_{\mu\nu} = \frac{q}{2I(2I-1)}\Phi_{\mu\nu}$, with $\Phi_{\mu\nu}$ denoting the the EFG tensor, $q$ as the scalar quadrupole moment of the nucleus, and $I$ as the nuclear spin. The gyromagnetic moment of the nucleus is written as $\gamma_n$. Note that we implicitly sum over double occurrences of Greek indices, a convention we keep throughout the manuscript. A detailed derivation of this Hamiltonian can be found in [13].

If the energy splitting due to the magnetic field is large compared to that caused by quadrupole interaction, the energy correction due to NQI can be treated perturbatively. In good approximation, the eigenstates of the total Hamiltonian can then be described by the eigenstates of the Zeeman-Hamiltonian, which are just the orientational nuclear spin states $|m_I\rangle$ for a fixed spin quantum number $I$. This leads to corrected transition energies

$$\Delta\mathcal{E}(m_I - 1 \to m_I) = -\gamma_n B_0 + \frac{3}{2}(2m_I - 1) Q_{zz}$$
$$\Delta\mathcal{E}(m_I - 2 \to m_I) = -2\gamma_n B_0 + \frac{3}{2}(4m_I - 4) Q_{zz}. \tag{2}$$

Note that this correction opens the possibility to address specific transitions individually, which is not possible in the case of equidistant Zeeman-splitting alone.

In order to drive transitions between different spin states $|m_I\rangle$ via quadrupole coupling, a time-dependent variation of the EFG tensor is necessary. Due to the quadratic nature of the NQI Hamiltonian, transitions with $\Delta m_I = \pm 1$ and $\Delta m_I = \pm 2$ can be driven. In a constant magnetic field $B = B_0 e_z$, the corresponding amplitudes are given by

$$g_{m_I \to m_I - 1}(t) = \alpha_{m_I - 1 \leftrightarrow m_I}(Q_{xz}(t) + iQ_{yz}(t));$$
$$\alpha_{m_I - 1 \leftrightarrow m_I} = \frac{1}{2}|2m_I - 1|\sqrt{I(I+1) - m_I(m_I - 1)}, \tag{3}$$

$$g_{m_I \to m_I - 2}(t) = \beta_{m_I - 2 \leftrightarrow m_I}(Q_{xx}(t) - Q_{yy}(t) + 2iQ_{yx}(t));$$
$$\beta_{m_I - 2 \leftrightarrow m_I} = \frac{1}{4}\sqrt{(I(I+1) - (m_I - 1)(m_I - 2))}$$
$$\times \sqrt{(I(I+1) - m_I(m_I - 1))}, \tag{4}$$

respectively. Note that the modulus of these transition amplitudes will determine the Rabi frequency of the respective spin level transitions in the dynamics simulations later.

The ONER protocol utilizes a periodic modulation of the steady-state occupation of an electronic two-level system to achieve a periodic modulation of the respective EFG tensor components. For a two-level system, in Born-Markov approximation, the steady-state density matrix is given by

$$\rho_{ee}(t \to \infty) = \frac{\Omega^2}{2\gamma_\perp \Gamma} \frac{1}{1 + \frac{\Delta^2}{\gamma_\perp^2} + \frac{\Omega^2}{\gamma_\perp \Gamma}}$$

$$\rho_{eg}(t \to \infty) = \frac{i\Omega}{2\gamma_\perp} \frac{1 + \frac{i\Delta}{\gamma_\perp}}{1 + \frac{\Delta^2}{\gamma_\perp^2} + \frac{\Omega^2}{\gamma_\perp \Gamma}}, \tag{5}$$

where $\Delta$ denotes the detuning of the laser frequency $\omega$ with respect to the transition energy of the two-level system $\omega_0$ from the ground state $|g\rangle$ to the excited state $|e\rangle$. The electronic Rabi frequency denoted as $\Omega$ depends on the intensity of the applied laser field and the dipole transition element of the two-level system. Parameters $\Gamma$ and $\gamma_c$ are the decay rate from excited state to ground state and the coherence decay rate, respectively. Note that we defined $\gamma_\perp = \frac{\Gamma}{2} + \gamma_c$.

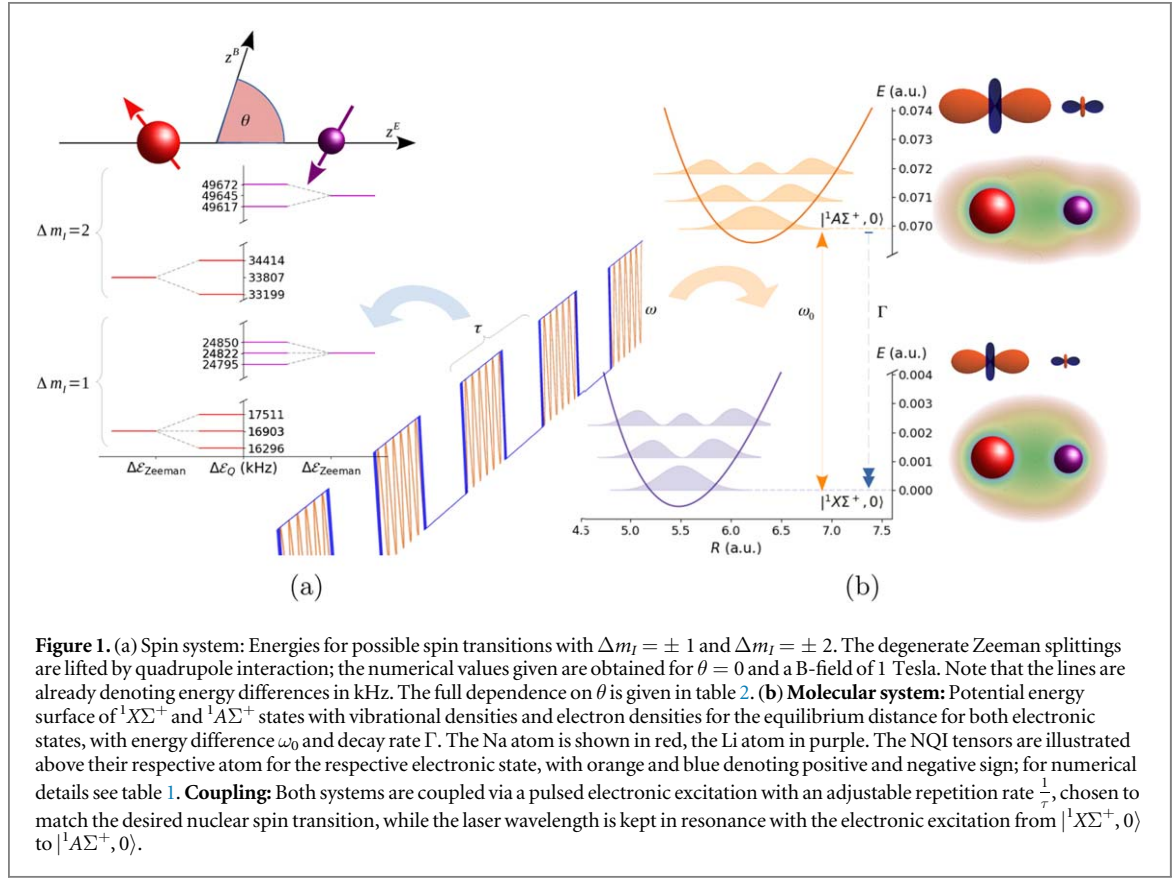

**Figure 1.** (a) Spin system: Energies for possible spin transitions with $\Delta m_I = \pm 1$ and $\Delta m_I = \pm 2$. The degenerate Zeeman splittings are lifted by quadrupole interaction; the numerical values given are obtained for $\theta = 0$ and a B-field of 1 Tesla. Note that the lines are already denoting energy differences in kHz. The full dependence on $\theta$ is given in table 2. (**b**) **Molecular system:** Potential energy surface of $^1X\Sigma^+$ and $^1A\Sigma^+$ states with vibrational densities and electron densities for the equilibrium distance for both electronic states, with energy difference $\omega_0$ and decay rate $\Gamma$. The Na atom is shown in red, the Li atom in purple. The NQI tensors are illustrated above their respective atom for the respective electronic state, with orange and blue denoting positive and negative sign; for numerical details see table 1. **Coupling:** Both systems are coupled via a pulsed electronic excitation with an adjustable repetition rate $\frac{1}{\tau}$, chosen to match the desired nuclear spin transition, while the laser wavelength is kept in resonance with the electronic excitation from $|^1X\Sigma^+, 0\rangle$ to $|^1A\Sigma^+, 0\rangle$.

## 3. System properties and computational details

We distinguish between 'molecular' and nuclear spin degrees of freedom. The energy splitting of the spin system is governed by Zeeman splitting in a constant external magnetic field $B$, defining the reference axis $z^B$. The reference axis for the molecular system is the symmetry axis of the molecule, which we denote as $z^E$; the latter can be oriented in principle by a constant, external electric field $E$ due to the dipole moment of the LiNa molecule. The angle between the two axes is denoted as $\theta$. Throughout the remaining manuscript, frame-dependent quantities are marked with superscripts $E$ or $B$ to indicate the respective frame. An illustration of the reference axes can be found in figure 1, our central graphics summarizing the whole concept of molecular ONER. Both systems, nuclear spin (figure 1(a)) and molecular system (figure 1(b)), are coupled via a sequence of laser pulses with suitable laser frequency $\omega$ and repetition rate $\frac{1}{\tau}$.

### 3.1. Molecular system
We begin with a discussion of relevant quantities of the molecular system. The first few electronic excitations of LiNa are in the orange to red visible spectrum of light, a range well covered e.g. by rhodamine-based dye or frequency-doubled Ti:sapphire lasers. For the calculation of electronic structure properties of the $^1X\Sigma^+$ ground state and the $^1A\Sigma^+$ electronically excited state of the singlet manifold we employ the Molpro program package, [14–16] combining a multiconfigurational self-consistent field method (MCSCF [17, 18]) with a follow-up multireference configuration interaction technique (MRCI [19, 20]) to account for effects of dynamic correlation. The $^1A\Sigma^+$ state has been chosen, since its equilibrium position is similar to the equilibrium of the ground state. Choosing $\Sigma$ states further avoids possible complications caused by a non-zero angular projection quantum number of the electron hull.

Table 1. Effective NQI tensor $\langle\phi|Q_{zz}^{(A),E}|\phi\rangle$ (kHz) for the two electronic states in their respective rovibrational ground state.

| $|\phi\rangle$ / A | Li | Na |
|---|---|---|
| $|^1X\Sigma^+, 0\rangle$ | −3.5 | 133.2 |
| $|^1A\Sigma^+, 0\rangle$ | −27.9 | 432.5 |

For reasons of computational convenience, the diatomic molecule is treated within the $C_{2v}$ molecular symmetry group. A balanced active space involving the orbitals 7/2/2/0 has been chosen with respect to the internal ordering $A_1/B_1/B_2/A_2$. The 1s orbital of Na is kept doubly occupied in all configurations. The aug-cc-pVTZ basis set [21] is used in all calculations.

We scan over the electronic energies as a function of internuclear distance and calculate the corresponding EFG tensor components for both electronic states. With this setup, very good agreement with the PES curves of [22] and [23] is achieved near the equilibrium geometries of both surfaces. On average, equilibrium distances for both states deviate by less than 4%, respectively. For the ground state, a distance of 5.48 bohr is obtained, for the excited state a distance of 6.22 bohr. PES scans are depicted in figure 1(b), alongside with their respective vibrational eigenstates and corresponding nuclear densities calculated numerically via finite differences on the respective PES. We obtain a vibrationally corrected transition energy $\omega_0$ between $|g\rangle := |^1X\Sigma^+, 0\rangle$ and $|e\rangle := |^1A\Sigma^+, 0\rangle$ of 459.87 THz (651.9 nm), for the vibrational ground states, and vibrational frequencies $\omega_{vib}$ of 7.46 and 6.35 THz (249 and 212 cm$^{-1}$) in both states, respectively. Even if monolithic, unstabilized solid-state laser systems with linewidths in the MHz regime are used in a future experimental realization, a single rovibrational line can be selected easily, and the molecular system can be regarded as an effective two-level system.

The mean NQI tensor $\langle\phi|Q_{\mu\nu}^{(A)}|\phi\rangle$ for each atom, i.e. $A \in \{Li, Na\}$, for the respective electronic state $\phi \in \{g, e\}$ is illustrated in figure 1(b) above the nuclear density plots, together with the electronic density distribution of the corresponding state. Numerical values are given in table 1. An assessment of the EFG prediction quality is possible through a comparison with known values for atomic Li and Na in their $^2P$ electronically excited states, falling back on the detailed studies of Sundholm and Olsen on atomic hyperfine parameters [24, 25]. Our EFG results (-0.02313 and -0.1098 a.u. for Li and Na, respectively) deviate from the corresponding non-relativistic limit by less than 2% on average. The magnitude of Sternheimer shielding effects lies within this uncertainty [26]; relativistic effects are fully negligible for the two smallest alkali metal atoms [27].

### 3.2. Nuclear spin system

The relevant parameters for the nuclear spin system are the gyromagnetic factor and the magnitude of the scalar NQI parameter. In the case of Li, both stable isotopes Li$^6$ and Li$^7$ show a non-zero scalar quadrupole moment of $-0.000806(6)$ barn and $-0.0400(3)$ barn, respectively [28, 29]. We choose Li$^7$ due to its larger value. The gyromagnetic moment of Li$^7$ is given by $\gamma_n^{Li7} = 3.256427(2) \mu_N \approx 24.8224 \frac{MHz}{T}$ as given in [29], where $\mu_N = 7.622593285(47) \frac{MHz}{T}$ is the nuclear magneton. For sodium, the only known stable isotope is Na$^{23}$, which features a significantly larger scalar quadrupole moment of $+0.104(1)$ barn. Its gyromagnetic moment is given by $\gamma_n^{Na23} = 2.217522(2) \mu_N \approx 16.9033 \frac{MHz}{T}$. Li$^7$ and Na$^{23}$ both have a total nuclear spin quantum number of $I = 3/2$, leading to an energy scale of several MHz for the Zeeman Hamiltonian of the spin system.

The mean NQI tensors of the states $|g\rangle$ and $|e\rangle$ for lithium and sodium are obtained via integration of the electronic expectation value of the NQI tensor, a function of the internuclear distance obtained from point-wise evaluations via Molpro, weighted by the corresponding nuclear probability density in the vibrational ground state. The results are cylindrical in the $E$-frame for both atoms, i.e. diagonal with equal $xx$ and $yy$ component, and since NQI tensors are also traceless, knowledge of e.g. the $zz$ component is sufficient. The vibrationally averaged magnitudes of the $zz$ components of the NQI tensor for the different electronic states are listed in table 1. The NQI tensor of sodium has a larger magnitude than that of lithium since its scalar quadrupole moment is larger by an order of magnitude.

Note that there is also spin-spin coupling between Li and Na, which is in the range of 2 kHz for the ground state and 1.6 kHz for the excited state, dominated by direct spin-spin coupling. This, however, does not interfere with the ONER protocol for single spin manipulation due to its small magnitude compared to Zeeman splitting, which mainly determines the repetition rate. Yet, its presence opens the possibility for spin-spin interaction, another pre-requisite for the future realization of qubit registers.

**Table 2.** Corrected spin transition energies and Rabi frequencies for spin transitions $m_I \leftrightarrow m_{I'}$ of $^7$Li and $^{23}$Na nuclear spin states, respectively, for optical excitation from $|g\rangle = |^1X\Sigma^+, 0\rangle$ to $|e\rangle = |^1A\Sigma^+, 0\rangle$. The corrected transition energies (necessary for the selection of the repetition rate $\frac{1}{\tau}$) and the Rabi frequencies are given as a function of the steady state population of the excited state $\rho_{ee}^\infty$ and the angle $\theta$ between system reference axis $z^E$ and spin laboratory frame $z^B$. The Zeeman splitting is given as a function of the external magnetic field $B$ measured in Tesla. The results are to be read as a single expression for each line, the compression of identical expressions in a column is just for better readability.

|        | $m_I$ | $m_{I'}$ | $\Delta\mathcal{E}(m_I \leftrightarrow m_{I'}|g,e)$ (kHz) | $|g(m_I \leftrightarrow m_{I'}|g,e)|$ (kHz) |
|--------|-------|----------|-----------------------------------------------------------|---------------------------------------------|
| $^7$Li  | $3/2 \leftrightarrow 1/2$   | | $24822.4\,B\;\mp$ | $\sin(2\theta)$ |
|        | $-1/2 \leftrightarrow -3/2$ | | $24822.4\,B\;\pm$ $\;(10.5 + 36.6\,\rho_{ee}^\infty) \times \frac{2 - 3\sin(\theta)^2}{2}$ | $20.2\,\rho_{ee}^\infty \times$ |
|        | $3/2 \leftrightarrow -1/2$  | | $49644.8\,B\;\mp$ | $\sin(\theta)^2$ |
|        | $1/2 \leftrightarrow -3/2$  | | $49644.8\,B\;\pm$ | |
| $^{23}$Na | $3/2 \leftrightarrow 1/2$ | | $16903.3\,B\;\pm$ | $\sin(2\theta)$ |
|        | $-1/2 \leftrightarrow -3/2$ | | $16903.3\,B\;\mp$ $\;(399.5 + 449.0\,\rho_{ee}^\infty) \times \frac{2 - 3\sin(\theta)^2}{2}$ | $247.6\,\rho_{ee}^\infty \times$ |
|        | $3/2 \leftrightarrow -1/2$  | | $33806.6\,B\;\pm$ | $\sin(\theta)^2$ |
|        | $1/2 \leftrightarrow -3/2$  | | $33806.6\,B\;\mp$ | |

### 3.3. ONER

Within the ONER protocol, we employ a pulsed excitation of the molecular two-level system $\{|g\rangle, |e\rangle\}$ to drive the nuclear spin transitions. Since the timescale of the latter is much larger than typical coherence times for electrons, an open electronic system has to be considered. For LiNa typical decay times of electronically excited states are in the range of ns, ie. $\Gamma \approx \mathcal{O}(\text{GHz})$. Thus, the system parameters satisfy the conditions

$$\omega \approx \omega_0 \gg \omega_{\text{vib}} \gg \Gamma \sim \Omega \gg |\gamma_n^{(A)} B_0| \gg \|Q\|, \tag{6}$$

where $\|Q\| := \max_{\mu\nu,A} |Q_{\mu\nu}^{(A)}|$ and $\Omega$ is the Rabi frequency of the molecular two-level transition. As shown in [13], the dynamical equations of the molecular system and the spin system can then be written as

$$\begin{aligned} i\partial_t \rho_M &\approx [H_M(t), \rho_M] \\ i\partial_t \rho_S &\approx \sum_A [(H_B^{(A)} + \langle Q_{\mu\nu}^{(A)}\rangle(t)\, I_\mu^{(A)} I_\nu^{(A)}), \rho_S] \end{aligned} \tag{7}$$

where $H_M(t)$ is the molecular (two-level) Hamiltonian in an external field with decay, $H_B^{(A)}$ denotes the Zeeman Hamiltonian of nucleus $A$, and the effective NQI tensor for nucleus $A$ is given as

$$\langle Q_{\mu\nu}^{(A)}\rangle(t) = \text{tr}_M\{\rho_M(t) Q_{\mu\nu}^{(A)}\} \stackrel{\text{RWA}}{\approx} \sum_i \rho_{i,i} \langle i|Q_{\mu\nu}^{(A)}|i\rangle, \tag{8}$$

where we applied the rotating wave approximation (RWA) for the non-diagonal elements of the effective NQI tensor, since electronic and vibrational motion take place at much shorter timescales than nuclear spin dynamics. This is generally true if the ONER conditions, equation (6), hold.

As discussed in [13], the relevant quantities for the ONER protocol are the constant NQI tensor $Q^{(0,A)}(e,g)$, which determines the quadrupole energy splitting (see equation (2)), and the harmonically modulated NQI tensor $Q^{(1,A)}(e,g)$, determining the magnitude of the Rabi frequency of the spin system (see equations (3) and (4)). Both can be expressed using the difference NQI tensor $\Delta Q^{(A)}(e,g) := \langle e|Q^{(A)}|e\rangle - \langle g|Q^{(A)}|g\rangle$ and the steady-state occupation of the excited state $\rho_{ee}^\infty$, yielding

$$Q^{(0,A)}(e,g) = \langle g|Q^{(A)}|g\rangle + \frac{\rho_{2L,ee}^\infty}{2}\Delta Q^{(A)}(e,g), \tag{9}$$

$$Q^{(1,A)}(e,g) = \frac{2\rho_{2L,ee}^\infty}{\pi}\Delta Q^{(A)}(e,g). \tag{10}$$

Both are easily obtained from the effective NQI tensor of the respective states in the $E$-frame given in table 1.

A transformation to the $B$-frame can be performed using the rotation matrix $R(\theta)$ around the shared axis of both frames. The respective NQI tensors in the $B$-frame are then given by

$$Q^{(j,A),B}(e,g) = R(\theta)\, Q^{(j,A),E}(e,g)\, R(\theta)^\top, \tag{11}$$

with $j \in \{0, 1\}$. The NQI tensor values in the $B$-frame can now be used to calculate the quadrupole energy correction and thus the correct repetition rate of the laser pulse, the $\Delta m_I = \pm 1$ and $\Delta m_I = \pm 2$ transition elements, and the corresponding Rabi frequency of the spin level transitions. In case of cylindrical NQI tensors, the $\theta$ dependence and magnitude can be separated; results for possible transitions of lithium and sodium are summarized in table 2. An illustration of the resulting transition energy splittings for $\theta = 0$ are shown in figure 1(a), assuming a B-field of 1 Tesla. As can be seen, all spin transitions on both atoms can be addressed individually, simply by adjusting the repetition rate $\frac{1}{\tau}$ of the laser pulses such that it matches the effective energy difference of the respective spin level transition. The effective energy difference is obtained by inserting the constant NQI tensor $Q^{(0,A),B}(e,g)$ into equation (2). The complete ONER scheme is illustrated in figure 1, where the spin system is coupled to the molecular system via a pulsed optical excitation, modulating the EFG tensor at the respective nuclei. Note that the $1/2 \to -1/2$ transition is not shown since it is quadrupole forbidden. In total, this leads to Rabi frequencies of the spin transitions with up to 10 kHz for Li and 120 kHz for Na. Depending on the choice of angle $\theta$, transitions with $\Delta m_I = \pm 1$ and $\Delta m_I = \pm 2$ can be driven. The former have their maximum at $\theta = \pi/4$, the latter at $\theta = \pi/2$. Note that transitions are only possible for $\theta \neq 0$, since a cylindrical NQI does not allow for spin level dynamics.

## 4. Conclusion

A possible realization pathway for the ONER protocol in molecules has been presented in theory, using LiNa as a diatomic benchmark system. Exploiting the $^1A\Sigma^+ \leftarrow {^1X\Sigma^+}$ electronic transition to modulate the electric field gradient at both nuclei, it should be possible to address various nuclear spin transitions either at the Li or the Na atom through a pulsed laser operating at 651.9 nm wavelength with a repetition rate between 16 and 50 MHz for a magnetic field of 1 Tesla, as tabulated in table 2. The underlying handle for this interaction is provided by the quadrupole moment of both nuclei, which couples to the electronic state-dependent electric field gradient. Molecular vibration is not compromising the protocol since it takes place on a much faster timescale than the nuclear spin transitions.

Based on this outcome, we believe that coherent, site-specific nuclear spin manipulation through well-established concepts of nonlinear optics and current laser technology is a realistic possibility. On the long run, more complicated spin manipulation tasks, as they are required for quantum computing, might become feasible, mapped into the time domain via pulse-duration encoded laser light. A logical next step is the investigation of optically controlled spin-spin coupling between specific nuclei in order to realize qubit registers through a combination of techniques.

## Acknowledgments

This research was funded in whole by the Austrian Science Fund (FWF) [10.55776/P36903]. For the purpose of open access, the author has applied a CC BY public copyright licence to any Author Accepted Manuscript version arising from this submission. Further support by NAWI Graz and the use of HPC resources provided by the IT services of Graz University of Technology (ZID) as well as the Vienna Scientific Cluster (VSC) is gratefully acknowledged.

## Data availability statement

All data that support the findings of this study are included within the article (and any supplementary files).

## ORCID iDs

Johannes K Krondorfer  https://orcid.org/0009-0009-5006-6319
Matthias Diez  https://orcid.org/0000-0001-6330-3082
Andreas W Hauser  https://orcid.org/0000-0001-6918-3106